\hfuzz 2pt
\font\titlefont=cmbx10 scaled\magstep1
\magnification=\magstep1

\null
\vskip 2cm
\centerline{\titlefont ON THE DECAY LAW}
\medskip
\centerline{\titlefont FOR UNSTABLE OPEN SYSTEMS}
\vskip 2.5cm
\centerline{\bf F. Benatti}
\smallskip
\centerline{Dipartimento di Fisica Teorica, Universit\`a di Trieste}
\centerline{Strada Costiera 11, 34014 Trieste, Italy}
\centerline{and}
\centerline{Istituto Nazionale di Fisica Nucleare, Sezione di Trieste}
\vskip 1cm
\centerline{\bf R. Floreanini}
\smallskip
\centerline{Istituto Nazionale di Fisica Nucleare, Sezione di Trieste}
\centerline{Dipartimento di Fisica Teorica, Universit\`a di Trieste}
\centerline{Strada Costiera 11, 34014 Trieste, Italy}
\vskip 2.5cm
\centerline{\bf Abstract}
\smallskip
\midinsert
\narrower\narrower\noindent
We use (nonconservative) dynamical semigroups to investigate the decay law
of a quantum unstable system weakly coupled with a large environment.
We find that the deviations from the classical exponential law are small
and can be safely ignored in any actual experiment.
\endinsert
\bigskip
\vfil\eject

The decay law of an unstable microscopic system can be rather well described
by an exponential function; this result can be easily justified on the
basis of classical probabilistic considerations. However, microscopic systems
should be described by quantum mechanics and it is well known that in quantum
theory the exponential decay law can not be valid for all times; in particular,
it surely fails for very short and very long times. Indeed, rather general
considerations assure that the quantum decay law can be described by a 
three-step function: a Gaussian law at short times, the classical exponential
law at intermediate times and finally a power law at longer times
({\it e.g.} see [1-3] and references therein).

Although various experiments have been devised in order to obtain evidence
for discrepancies with the exponential law, none have been so far actually
detected.[4] In this respect, a system that has attracted a lot of interest
both theoretically and experimentally is the neutral kaon system.
The $K^0$-$\overline{K^0}$ system has proven to be one of the most fruitful
systems for testing fundamental symmetries, like $CP$ and $CPT$.
In parametrizing violations of these symmetries, one usually takes for
granted the exponential decay law and uses an effective theory to describe
the kaon system.[5]

In the following we shall examine to what extent the exponential decay law
can be trusted in actual experiments. More precisely, we shall study the
deviations from the exponential law for small and large times, taking also into
account possible effects due to incoherent interactions with the environment.

The general idea that is at the basis of our considerations is that unstable
systems can be viewed as specific examples of open quantum systems.
These systems can be modeled in general as being small subsystems in weak
interaction with large environments. Although the global time evolution
of the closed compound system is described by an unitary transformation,
the reduced dynamics of the subsystem, obtained by the elimination of the
environment degrees of freedom, usually develops some sort of dissipation and
irreversibility. Under mild assumptions, the reduced evolution is realized by
one-parameter (=time) maps acting on the states of the system, conveniently
described by density matrices, with forward in time composition (semigroup
property) and the additional characteristic of being completely positive.
This set of transformations forms a so-called dynamical semigroup.[6-10]

This rather universal and general formalism has been recently adopted to
treat effective dynamics for the kaon system that transform pure states into
mixed ones.[11-16] The physical motivations behind such an approach are based on
quantum gravity, that predicts loss of quantum coherence at the Planck's
length due to fluctuations of the gravitational field.[17] These generalized
time-evolutions lead to $CP$ and $CPT$ violating effects that could be in
the reach of the next generation of neutral kaon experiments.[15, 16]

In these treatments, the environment was assumed not to contribute to the decay
of the kaons, that was effectively described by the standard exponential law.
However, possible effects of the environment on the decay process itself
are surely conceivable. As we shall see, they can be studied using again
the formalism of dynamical semigroups. It will turn out that these
environment effects do not modify the short-time behaviour of the decay law,
but only the exponential and power law regimes. 

However, in the realistic hypothesis of a weak coupling between 
unstable system and environment, these modifications are tiny or occur for 
too large times for any practical considerations. 
Therefore, the exponential decay law captures rather well the essential features
of the decay process of an unstable system and can be certainly used with
confidence in any actual experiment involving weak-decaying particles,
like the neutral $K$ mesons.

\bigskip

For our considerations, we choose to describe the states of a quantum system
evolving in time by means of density matrices $\rho(t)$. Given an initial
state $\rho$ and a time-independent hermitian hamiltonian $H$, 
the standard time evolution is described by
$$
\rho(t)= e^{-iHt}\ \rho\ e^{iHt}\ ,\eqno(1)
$$
solution of the Liouville-von Neumann equation
$$
{\partial\rho(t)\over\partial t}=-i[H,\rho(t)]\ .\eqno(2)
$$
In the case of an unstable system, it is custom to split $H$ as
$$
H=H^{(0)}+H^{(1)}\ ,\eqno(3)
$$
where $H^{(0)}$ is the unperturbed hamiltonian, while $H^{(1)}$ is the
interaction hamiltonian that drives the decay process; the system
would be stable if $H^{(1)}=\,0$.
In the case of the neutral $K$-mesons, $H^{(0)}$
can be identified with the hamiltonian of the strong interactions and
$H^{(1)}$ with that of the weak interactions (possible ``superweak''
mixing terms should be included in $H^{(0)}$). Further, we shall call $P_u$
the projector operator on the subspace of the undecayed states. We shall
also use the orthogonal projector $P_d=1-P_u$; it describes the transition
to the space of the decayed states.

Assuming that at the beginning our unstable system is in the undecayed state
$\rho=P_u\,\rho\, P_u$, the probability of finding it undecayed at time $t$
is given by
$$
{\cal P}(t)={\rm Tr}[\rho(t)\, P_u]\ .\eqno(4)
$$
In the case of the decay of a single particle, with evolution as in (1),
the properties of ${\cal P}(t)$ have been widely studied in the literature.
Here we shall extend those treatments by considering also the interaction
of the decay system (not necessarily one-dimensional) with the environment.
Together with the effects of the hamiltonian $H^{(1)}$, also this interaction
could in principle contribute to the decay process.

As mentioned in the introductory remarks, in order to take into account these
extra effects we shall treat the unstable system 
as an open quantum system $S$.[6-10]
As for any open system, $S$ can be thought of as interacting with a suitable 
environment $E$, so that the global system $S+E$ is closed. This evolves
in time according to a group of unitary operators as in (1), governed by a
total hamiltonian $H_{S+E}$, which is the sum of the hamiltonian $H_S$
of the subsystem, of the hamiltonian $H_E$ of the environment and of the
interaction hamiltonian $H_{SE}$ between them. A reduced dynamics for the
subsystem $S$ can be consistently obtained when $S$ and $E$ are assumed
to be uncorrelated at the moment of the formation of the unstable
system and therefore the state of the total system is simply: 
$\rho_{S+E}=\rho\otimes\rho_E$. 

In such cases, by tracing over the the environment degrees of freedom
one gets linear, completely positive maps
$\rho \mapsto \tilde{\gamma}_t[\rho]$ on the states of the subsystem, where
$$
\tilde{\gamma}_t[\rho]\equiv
{\rm Tr}_E\Big[e^{-iH_{S+E}t}\big(\rho\otimes\rho_E\big)
e^{iH_{S+E}t}\Big]\ .\eqno(5)
$$
These transformations do not have any 
simple composition law and they usually
contain memory effects.
However, when the interaction with the environment is weak, they reduce
to completely positive, dissipative dynamical maps
$\gamma_t: \rho\mapsto\rho(t)$, which obey a semigroup composition law, 
$\gamma_{t_1}\circ\gamma_{t_2}=\gamma_{t_1+t_2}$,
$t_1,t_2\geq 0$, and satisfy the condition of entropy increase:
$dS/dt\geq0$, $S(t)=-{\rm Tr}[\rho(t)\ln\rho(t)]$. 
Moreover, these semigroups are generated by equations of
a very specific type that can be explicitly given:[6]
$$
{\partial\rho(t)\over\partial t}=R\, \rho(t)+\rho(t)\, R^\dagger +
\sum_k A_k\, \rho(t)\, A_k^\dagger\ ,\eqno(6)
$$
where the set of (bounded) operators $R$ and $A_k$ are such that:
$R+R^\dagger+\sum_k A_k^\dagger A_k\leq0$. They are probability
preserving or not depending on whether this combination
is identically zero or not; they are called quantum dynamical semigroups.
Notice that this description of open systems is rather general and the
effective subdynamics $\gamma_t$ is essentially independent from the type
of the environment. 

In the case of unstable high energy particle systems, a natural choice for $E$
could be the gravitational field, whose effects are usually neglected
in the theory of elementary particles because of the smallness of its coupling.
Its quantum fluctuations at Planck's length could nevertheless act as
a weak coupled environment producing detectable effects in elementary particle
interactions.[17, 11] It is also worth mentioning 
that the effects of such fluctuations,
realized via the space-time foam, can be effectively described with an heat
bath, the most natural of all environments.[18]

In view of all above considerations, we shall now generalize the standard 
quantum mechanical evolution equations for an unstable system by adding
to (2) a linear piece of the form given by the r.h.s. of (6). Since this
additional piece should describe the contribution of the environment to
the decay process, it must be proportional to $\rho(t)-P_u\,\rho(t)\,P_u$.
In fact, this term would project $\rho(t)$ out of the space of
undecayed states. Thus, we shall study the generalized time evolution
described by the equation
$$
{\partial\rho(t)\over\partial t}=-i\big[ H,\rho(t)\big]
-\lambda\big(\rho(t)-P_u\,\rho(t)\, P_u\big)\ ,\eqno(7)
$$
with $\lambda$ a positive ``coupling'' constant and $H$ as in (3).
We shall consider weak coupled environments, and therefore 
assume $\lambda$ to be much smaller than any typical energy scale
in the hamiltonian. In the case of the $K^0$-$\overline{K^0}$ system,
assuming the dissipative term in (7) of gravitational origin,
dimensional arguments suggest $\lambda$ to be at most of order
$m_K^2/m_P$, where $m_K$ is the kaon rest mass and $m_P$ is Planck's mass.
The evolution equation (7) describes nonconservative dynamical maps.[6] 
Indeed, probability is not conserved,
$(d/dt){\rm Tr}[\rho(t)]=-\lambda {\rm Tr}[\rho(t)\, P_d]\leq0$,
as it should be for an unstable system; however, this violation is small, since
$\lambda$ is small.

Equations of the form (7) were used before in discussing one-dimensional
unstable systems.[1]
The motivation for the introduction of the nonstandard term 
was there attributed to the interaction of the decaying system with
the measuring apparatus. While this point of view is surely viable, we stress
that our interpretation of (7) as describing the evolution of an unstable system
is completely phenomenological in nature.
In particular, we do not make any specific assumption on the phenomena
responsible for the appearance of the second term in the r.h.s. of (7).

In order to proceed in the study of equation (7), it is convenient to introduce
a vector notation, rewriting the density matrix $\rho$ describing the state
of the unstable system as the vector $|\rho\rangle$. To any operator acting
on the state $\rho$, one can define a corresponding operator acting on the vector
$|\rho\rangle$. In particular, one can define the projector ${\mit\Pi}_u$,
$$
{\mit\Pi}_u |\rho\rangle\equiv \big|P_u\, \rho\, P_u\big\rangle\ ,\qquad
{\mit\Pi}_u^2={\mit\Pi}_u\ ,\eqno(8)
$$
and its orthogonal complement: ${\mit\Pi}_d=1-{\mit\Pi}_u$. 
By introducing the Liouville operator
$L_H$ corresponding to the hamiltonian $H$,
$$
L_H\ |\rho\rangle\equiv\big|\, [H,\rho]\,\big\rangle\ ,\eqno(9)
$$
one can rewrite (7) as a Schr\"odnger like equation:
$$
i{\partial\over\partial t}|\rho(t)\rangle= L\ |\rho(t)\rangle\ ,\eqno(10)
$$
with
$$
L\equiv L_H-i\lambda\ {\mit\Pi}_d\ .\eqno(11)
$$
Although the statements and the conclusions obtained below using this new
formalism can be rigorously justified, for sake of simplicity we shall keep
mathematical considerations to a minimum. As we shall see in the following,
this formalism results particularly appropriate in the study of unstable
system for which the space of the undecayed states is not one-dimensional,
as in the case of the neutral kaons.

We are ultimately interested in describing the properties of the probability
${\cal P}(t)$ in (4). Therefore, one should concentrate on the
study of the time evolution of the projected vector:
$$
|\rho(t)\rangle_u={\mit\Pi}_u\ |\rho(t)\rangle\ .\eqno(12)
$$
Using the Laplace transformed vector $|\tilde{\rho}(s)\rangle$, the 
corresponding Lippmann-Schwinger equation reads:
$$
\bigg[s+i\bigg(L_{uu}+L_{ud}\ {1\over is-L_{dd}+i\lambda{\mit\Pi}_d}\ L_{du}
\bigg)\bigg]\ |\tilde{\rho}(s)\rangle_u=|\tilde\rho(0)\rangle_u\ , \eqno(13)
$$
where $s$ is the Laplace variable and
$$
L_{uu}={\mit\Pi}_u L_H {\mit\Pi}_u\ ,\quad
L_{ud}={\mit\Pi}_u L_H {\mit\Pi}_d\ ,\quad
L_{du}={\mit\Pi}_d L_H {\mit\Pi}_u\ ,\quad
L_{dd}={\mit\Pi}_d L_H {\mit\Pi}_d\ . \eqno(14)
$$
The action of these operators on any state $|\rho\rangle$ is well-defined.
For instance, using the shorthand notation ${\cal O}_{ij}=P_i\, {\cal O}\, P_j$,
$i,j=u,d$, with $\cal O$ a generic operator, one finds:
$$
\eqalignno{
&L_{ud}\ |\rho\rangle=\big| H_{ud}\, \rho_{du}-\rho_{ud}\, H_{du}\big\rangle\ ,
&(15a) \cr
&L_{du}\ |\rho\rangle=\big| H_{du}\, \rho_{uu}-\rho_{uu}\, H_{ud}\big\rangle\ .
&(15b) }
$$

According to our hypothesis, the decay process is driven by the interaction
hamiltonian $H^{(1)}$ and the coupling to the environment. In particular,
the decaying system would be actually stable for $H^{(1)}=\, 0$ and
$\lambda=\,0$. This immediately implies that $H_{ud}\equiv H_{ud}^{(1)}$,
$H_{du}\equiv H_{du}^{(1)}$ and $H_{uu}\equiv H_{uu}^{(0)}$. 
Furthermore, as $H^{(1)}$ is supposed to be
small (in the case of the neutral kaons, $H^{(1)}$ is indeed the weak
hamiltonian), we shall study approximated solutions of (13), taking into
account only terms up to second order in $H^{(1)}$; this is also the
approximation that is usually adopted in the description of decaying
particles in standard quantum mechanics.

At this point, in order to simplify the formulas, we also assume that the space
of the undecayed states is degenerate in energy:
$H_{uu}\equiv H_{uu}^{(0)}=E_0\, P_u$. In the case of the neutral kaons,
this is not really a restriction since $K^0$ and $\overline{K^0}$ have
the same rest mass. Then, by acting on a generic state $|\rho\rangle_u$,
one can prove that:
$$
L_{uu}+L_{ud}\ {1\over i(s+\lambda){\mit \Pi}_d-L_{dd}}\ L_{du}=
{\cal L}_{W(s)}+O\Big((H^{(1)})^3\Big)\ , \eqno(16)
$$
where $W(s)$ is the effective non-hermitian ``hamiltonian''
$$
W(s)=H_{ud}^{(1)}\ {1\over i(s+\lambda)+E_0-H_{dd}^{(0)}}\
H_{du}^{(1)}\ ,\eqno(17)
$$
and ${\cal L}_{W(s)}$ is the corresponding generalized Liouville operator
$$
{\cal L}_{W(s)}|\rho\rangle = \Big|W(s)\,\rho-\rho\, W^\dagger(s)\Big\rangle\ .
\eqno(18)
$$
This is not surprising since, in view of (15), the generalized operators
$L_{ud}$ and $L_{du}$ themselves are of order $H^{(1)}$.
Thus, up to second order terms in $H^{(1)}$, equation (13) becomes
$$
\Big(s+i{\cal L}_{W(s)}\Big)\ |\tilde{\rho}(s)\rangle_u = 
|\tilde\rho(0)\rangle_u\ .\eqno(19)
$$
Using the inverse Laplace transform, one can then write
$$
|\rho(t)\rangle_u={1\over 2\pi i}\int_{c-i\infty}^{c+i\infty}
ds\ e^{st}\ \bigg[{1\over s+i {\cal L}_{W(s)}}\bigg]\ 
|\tilde\rho(0)\rangle_u\ ,\eqno(20)
$$
where $c$ must be chosen so that the integration path in the complex
$s$-plane lies to the right of all singularities of the generalized
operator $[s+i{\cal L}_{W(s)}]^{-1}$.

Before proceeding further, let us first deduce from (20) the standard
exponential decay law. The singularities of the integrand in (20) are related
to those of the Liouville operator ${\cal L}_{W(s)}$, and therefore to
those of $W(s)$ in (17). Indeed, using the definition (18), one can deduce
that the spectrum of ${\cal L}_{W(s)}$ coincides with the difference
of the spectra of $W(s)$ and $W^\dagger(s)$. The operator $W(s)$ is analytic
in the complex $s$-plane except when the denominator in (17) vanishes. 
Since typically the spectrum of $H_{dd}^{(0)}$ is continuum, the singularities
of $W(s)$ lay on the imaginary axis, where usually there is a cut.
Therefore, one can conclude that also $[s+i{\cal L}_{W(s)}]^{-1}$ has
generally a cut on the imaginary axis of the $s$-plane.
One can then move the integration path in (20)
almost to coincide with this axis. 

Further, one notice that when
$H^{(1)}=\,0$ and $\lambda=\,0$ the generalized
Liouville operator ${\cal L}_{W(s)}$ becomes the
null operator, so that only a pole at the origin occurs in the 
integrand of (20).
In this case, we obtain: $|\rho(t)\rangle_u=|\rho(0)\rangle_u$, as it should be,
since now the system is stable. 
In presence of interactions, this pole moves into the
second sheet, but remains close to the imaginary axis for small $H^{(1)}$
and $\lambda$.[3] One then expects that this pole continues to give the main
contribution to the integral. This is the so called Weisskopf-Wigner
approximation, that gives rise to the exponential decay law for all times.
Indeed, within this approximation and using (18), the integral in (20) gives
$$
|\rho(t)\rangle_u=e^{-it{\cal L}_{H_W}}\ |\rho(0)\rangle_u\ ,\eqno(21)
$$
or equivalently
$$
\rho_{uu}(t)=e^{-iH_Wt}\ \rho_{uu}(0)\ e^{iH_W^\dagger t}\ ,\eqno(22)
$$
where the effective hamiltonian $H_W$ takes the form:
$$
H_W=H_{uu}^{(0)}+H_{ud}^{(1)}\ {1\over E_0-H_{dd}^{(0)}+i\lambda}\
H_{du}^{(1)}\ .\eqno(23)
$$
This hamiltonian is not hermitian and can be written as $H_W=M-i{\mit\Gamma}/2$,
with $M$ and $\mit\Gamma$ hermitian and positive. Indeed, by putting the system
in a finite box so that the spectrum of $H_{dd}^{(0)}$ becomes discrete,
one explicitly finds:
$$
\eqalignno{
&[M]_{\alpha\beta}=E_0\, \delta_{\alpha\beta}+
\sum_k\langle\alpha|H^{(1)}|k\rangle\ {(E_0-E_k)\over (E_0-E_k)^2+\lambda^2}\
\langle k| H^{(1)}|\beta\rangle\ , &(24a)\cr
&[{\mit\Gamma}]_{\alpha\beta}=2 \sum_k\langle\alpha|H^{(1)}|k\rangle\ 
{\lambda\over (E_0-E_k)^2+\lambda^2}\
\langle k| H^{(1)}|\beta\rangle\ , &(24b)}
$$
where we have used greek (latin) indices to label undecayed (decay-products) 
states. The entries of the matrix
$$
[\sigma(E)]_{\alpha\beta}=\sum_{E_k\leq E} \langle\alpha|H^{(1)}|k\rangle\ 
\langle k| H^{(1)}|\beta\rangle\ ,\eqno(25)
$$
are usually found to be piecewise differentiable in 
the variable $E$ when the volume of the box
becomes infinite.[2] In this limit, by setting
$$
[\omega(E)]_{\alpha\beta}={d[\sigma(E)]_{\alpha\beta}\over dE}\ ,\eqno(26)
$$
the sums in (24) can be substituted by integrals:
$$
\eqalignno{
&[M]_{\alpha\beta}=E_0\, \delta_{\alpha\beta}
+ \int_{E_m}^\infty dE\ [\omega(E)]_{\alpha\beta}\, 
{(E_0-E)\over (E_0-E)^2+\lambda^2}\ , &(27a)\cr
&[{\mit\Gamma}]_{\alpha\beta}=2\int_{E_m}^\infty dE\ 
[\omega(E)]_{\alpha\beta}\, 
{\lambda\over (E_0-E)^2+\lambda^2}\ , &(27b) }
$$
where $E_m$ is the lowest eigenvalue of $H_{dd}^{(0)}$.
For $\lambda$ small, using
$$
\eqalignno{
&{x\over x^2+\lambda^2}=P{1\over x}+\lambda\,\pi\delta'(x)+O(\lambda^2)
\ , &(28a)\cr
&{\lambda\over x^2+\lambda^2}=\pi\delta(x)+\lambda\, P{1\over x^2}+
O(\lambda^2)\ ,&(28b) }
$$
with $P$ indicating principal value, the equations (27) finally become:
$$
\eqalignno{
&[M]_{\alpha\beta}=[\mu(E_0)]_{\alpha\beta} + \lambda\pi\, 
[\omega'(E_0)]_{\alpha\beta}\ , &(29a)\cr
&[{\mit\Gamma}]_{\alpha\beta}=2\pi\, [\omega(E_0)]_{\alpha\beta} +
2\lambda\, \big(\delta_{\alpha\beta}-[\mu'(E_0)]_{\alpha\beta}\big)\ , &(29b) }
$$
where the dash signifies derivative with respect to $E_0$ and
$$
[\mu(E_0)]_{\alpha\beta}=E_0\, \delta_{\alpha\beta} +
\int_{E_m}^\infty dE\ [\omega(E)]_{\alpha\beta}\ P{1\over E_0-E}\ .
\eqno(30)
$$
When $\lambda=\,0$, $M$ and $\mit\Gamma$ reduce to their standard
Wiesskopf-Wigner expressions. The effect of the environment is to
modify these expressions by adding terms that are in principle calculable using
field theory techniques.[19-21] 
The actual evaluation of the various terms in (29)
requires the adoption of a specific microscopic model
for the interacting hamiltonian and is certainly beyond the purpose of the
present work.

As stressed at the beginning, the exponential decay law can not hold for all times.
We shall now go back to the equation (19), which is exact up to second
order terms in $H^{(1)}$, and try to evaluate the corrections to the
Weisskopf-Wigner approximation. To this purpose, let us add and subtract
to the l.h.s. of (19) the term $i{\cal L}_{H_W}$.[22] Within our approximation,
one can then write:
$$
\Big[s+i{\cal L}_{W(s)}\Big]^{-1}=\Big[s+i{\cal L}_{H_W}\Big]^{-1}
-{i\over s^2}\Big({\cal L}_{W(s)}-{\cal L}_{H_W}\Big)\ .\eqno(31)
$$
By acting on a generic state $|\rho\rangle_u$, using simple algebra
one can prove that:
$$
-{i\over s^2}\Big({\cal L}_{W(s)}-{\cal L}_{H_W}\Big)\ |\rho\rangle_u 
=\big|\, V(s)\,\rho_{uu} - \rho_{uu}\, V^\dagger(s)\,\big\rangle\ ,\eqno(32)
$$
where
$$
V(s)=H_{ud}^{(1)}\ {1\over (E_0-H_{dd}^{(0)}+i\lambda)^2}\
\Bigg[ {1\over i(s+\lambda)+E_0-H_{dd}^{(0)}}-{1\over is}\Bigg]\ 
H_{du}^{(1)}\ .\eqno(33)
$$
Inserting these results in (20) and performing the $s$-integration, one obtains
an effective evolution for the projected density matrix $\rho_{uu}$ of the
form
$$
\rho_{uu}(t)= U(t)\ \rho_{uu}(0)\ U^\dagger(t)\ ,\eqno(34)
$$
where
$$
U(t)=e^{-iH_Wt}+H_{ud}^{(1)}\ {1\over (E_0-H_{dd}^{(0)}+i\lambda)^2}\
\Big[ e^{-i(H_{dd}^{(0)}-i\lambda)t}-e^{-iE_0 t}\Big]\ 
H_{du}^{(1)}\ .\eqno(35)
$$
This evolution satisfies the correct boundary condition, $U(0)=P_u$;
it contains contributions up to second order in $H^{(1)}$
and to all orders in $\lambda$.
It is dominated by the exponential Weisskopf-Wigner term; however, the additional
pieces are relevant for small and large times.

By expanding (35) for small $t$, one gets:
$$
U(t)\ \sim_{_{\hskip -10pt t\rightarrow0}}\ (1-it E_0) P_u-{t^2\over2}
\Big( E_0^2 P_u+ H_{ud}^{(1)}\, H_{du}^{(1)}\Big)\ .\eqno(36)
$$
Inserting this result in (34), the probability (4) takes the form:
$$
{\cal P}(t)\simeq 1-(t/\tau_G)^2\equiv
1-t^2\ {\rm Tr}\Big[H_{ud}^{(1)}\, H_{du}^{(1)}\ \rho_{uu}(0)\Big]\ ,\eqno(37)
$$
and therefore has a Gaussian behaviour;
the Gaussian width $\tau_G^{-2}$, that can be rewritten as
${\rm Tr}\big[H_{ud}\, H_{du}\ \rho_{uu}(0)\big]$, gives the 
spread in energy $(\Delta E)^2$ of the initial state $\rho_{uu}(0)$.
Furthermore, ${\cal P}(t)$ is independent
from $\lambda$, so that the small-time decay law is unaffected by the
interaction with the environment. This result is physically understandable:
the time is too short to allow the environment to play a role in the decay,
which, in this early stages, is totally driven by $H^{(1)}$. This is also
in agreement with our starting assumption that the environment does not
disturb the preparation of the decaying system.

The times for which the Gaussian behaviour of ${\cal P}(t)$ can be, at least
in principle, experimentally detected are in general very small. A comparison
with the exponential behaviour in (22), which gives
$$
{\cal P}(t)\simeq 1-t/\tau\equiv
1-t\ {\rm Tr}\big[{\mit\Gamma}\, \rho_{uu}(0)\big]\ ,\eqno(38)
$$
indicates that this could happen only for times smaller than 
$t_m\simeq \tau_G^2/\tau=1/(\Delta E)^2\tau$.
However, by taking into account the Heisenberg
time-energy uncertainty principle, $\Delta E\, \Delta t\geq 1$, one
finds $t_m$ to be actually smaller than
the time interval $\Delta t$ necessary to complete any measurement.[1, 23] 
Therefore, it is practically impossible to detect 
deviations from the standard exponential
decay law at small times in any actual experiment involving elementary
particles (nevertheless, in suitable atomic systems the situation 
might be different, see [24, 25]).

In studying the large time behaviour of $U(t)$, it is convenient to rewrite
(35) as:
$$
U(t)=\big[ 1-R(0)\big]\ e^{-iH_W t} +R(t)\ ,\eqno(39)
$$
where
$$
R(t)=H_{ud}^{(1)}\ {1\over (E_0-H_{dd}^{(0)}+i\lambda)^2}\
e^{-i(H_{dd}^{(0)}-i\lambda)t}\ H_{du}^{(1)}\ .\eqno(40)
$$
For large times, the prefactor to the Wiesskopf-Wigner term, which is
actually equal to $dH_W(E_0)/dE_0$, can be reabsorbed in a normalization
of the states. By putting the system in a box and using again the definitions
(25) and (26), in the infinite-volume limit the correction term $R(t)$ takes the
form
$$
R(t)=\int_{E_m}^\infty dE\ \omega(E)\ {e^{-i(E-i\lambda)t}\over
(E_0-E+i\lambda)^2}\ .\eqno(41)
$$
Due to phase space limitations, the behaviour on threshold of $\omega(E)$
can be usually approximated by:
$$
\omega(E)\simeq\omega(E_0)\bigg({E\over E_m}-1\bigg)^\delta\ ,\qquad
\delta>0\ .\eqno(42)
$$
Inserting this expression in (41), together with a convergent factor
$e^{-\varepsilon E}$, $\varepsilon\ll t$, the integral in $R(t)$ can be
evaluated exactly in terms of Whittaker functions.[26] 
Taking the large $t$ limit, one finally obtains
$$
R(t)\, \sim_{_{\hskip -12pt t\rightarrow\infty}}\, 
\Gamma(\delta+1)\ {E_m\ \omega(E_0)\over (E_0-E_m+i\lambda)^2}\
{e^{-i(E_m-i\lambda)t}\over\big(it\, E_m\big)^{\delta+1} }\ .\eqno(43)
$$
Therefore, for large times the effective evolution operator $U(t)$ exhibits
a power law behaviour, modulated by an exponential.[22]
Notice that the probability ${\cal P}(t)$ in (4) has a more 
complicated behaviour
due to the interference effects between the two terms in (39). In fact,
besides the standard Wiesskopf-Wigner term proportional to
$e^{-t/\tau}$ and the power-like 
term proportional to $t^{-2(\delta+1)}$,
${\cal P}(t)$ contains also an oscillating term that, taking for simplicity
$\lambda=\,0$, is proportional to 
$e^{-t/2\tau}\, {\cal R}e[e^{-itE_m}/(it)^{\delta+1}]$.
In another context, this intermediate behaviour has also been noticed in [24].

In order to estimate the region in which 
the long-time correction (43) supersedes
the standard exponential decay term, one has to consider the magnitude
of the correction, given by the modulus $|R(t)|$ of the operator in (43).
Recalling $(29b)$, one has
$$
|R(t)|\simeq \big|{\mit\Gamma}-2\lambda\, {\mit\Delta\Gamma}\big|\
{E_m\over (E_0-E_m)^2+\lambda^2}\
{e^{-\lambda t}\over (t\, E_m)^{\delta+1} }\ ,\eqno(44)
$$
where ${\mit\Delta\Gamma}=1-\mu'(E_0)$. This result should be
compared with the standard exponential term $e^{-{\mit\Gamma}t/2}$.
In order to do this, let us choose a basis in which $\mit\Gamma$ (or better
the spectral operator $\omega$) is diagonal, and label with the
index $\alpha$ the corresponding eigenvalues. In the case of the
neutral kaon system, $\alpha$ takes the two values $S$ and $L$, which
refer to the $K_S$ and $K_L$ states.
Assuming the difference $E_0-E_m$ of the same order of $E_m$,
one finds that the power law behaviour dominates for times larger 
than $\tau_\alpha/{\mit\Gamma}_\alpha$, where $\tau_\alpha$
is implicitly given by the following equation
$$
\tau_\alpha=2\, \bigg(1+2{\lambda\over{\mit\Gamma}_\alpha}\bigg)\, 
(\delta+1)\ln\tau_\alpha 
+ 2\, \bigg(1+2{\lambda\over{\mit\Gamma}_\alpha}\bigg)\, (\delta+2) 
\ln\bigg({E_m\over{\mit\Gamma}_\alpha}\bigg)
+2\lambda {\mit\Delta\Gamma_\alpha\over\Gamma_\alpha}\ .\eqno(45)
$$
Since $E_m/{\mit\Gamma}_\alpha$ is usually very large 
$(\sim 10^{15})$ in elementary
particle decays, it turns out that deviations from the exponential law can be
seen only after hundreds of life-times (for $\delta=1$ and $\lambda=\,0$),
a region which is clearly unavailable to the experiment. Furthermore,
notice that the role of the environment tends to worsen the situation, 
pushing the limit towards even longer times.

We can safely conclude that the role played by the environment in the decay
process of an unstable system is marginal and can not be detected in actual
experiments. This conclusion results from a careful study of the starting
evolution equation (7). Although phenomenological in nature,
this equation encodes in a rather general and universal way possible
dissipative effects due to a weak interaction with an environment.
From this point of view, the description of the neutral
kaon system in terms of completely positive dynamical semigroups given
in [13-16] is appropriate. In particular, the deviations from the predictions
of the standard Weisskopf-Wigner theory discussed there, if experimentally
detected, could be really the sign of new physics.

\vskip 2cm

\centerline{\bf ACKNOWLEDGMENTS}
\bigskip

We thank N. Paver for many illuminating discussions.

\vfill\eject

\centerline{\bf REFERENCES}
\bigskip

\item{1.} L. Fonda, G.C. Ghirardi and A. Rimini, Rep. Prog. Phys.
{\bf 41} (1978) 587
\smallskip
\item{2.} A. Peres, Ann. of Phys. {\bf 129} (1980) 33
\smallskip
\item{3.} H. Nakazato, M. Namiki and S. Pascazio, Int J. Mod. Phys. {\bf B10}
(1996) 247
\smallskip
\item{4.} N.N. Nikolaev, Sov. Phys. Usp. {\bf 11} (1968) 522; for more
recent results, see: The OPAL Collaboration, Phys. Lett. {\bf B368} (1996) 244
\smallskip
\item{5.} T.D. Lee and C.S. Wu, Ann. Rev. Nucl. Sci. {\bf 16} (1966) 511
\smallskip
\item{6.} R. Alicki and K. Lendi, {\it Quantum Dynamical Semigroups and 
Applications}, Lect. Notes Phys. {\bf 286}, (Springer-Verlag, Berlin, 1987)
\smallskip
\item{7.} H. Spohn, Rev. Mod. Phys. {\bf 53} (1980) 569
\smallskip
\item{8.} G. Lindblad,  Comm. Math. Phys. {\bf 48} (1976) 119
\smallskip
\item{9.} E.B. Davies, {\it Quantum Theory of Open Systems}, (Academic Press,
New York, 1976)
\smallskip
\item{10.} V. Gorini, A. Frigerio, M. Verri, A. Kossakowski and
E.C.G. Surdarshan, Rep. Math. Phys. {\bf 13} (1978) 149 
\smallskip
\item{11.} J. Ellis, J.S. Hagelin, D.V. Nanopoulos and M. Srednicki,
Nucl. Phys. {\bf B241} (1984) 381; J. Ellis, J.L. Lopez, N.E. Mavromatos 
and D.V. Nanopoulos, Phys. Rev. D {\bf 53} (1996) 3846
\smallskip
\item{12.} P. Huet and M.E. Peskin, Nucl. Phys. {\bf B434} (1995) 3
\smallskip
\item{13.} F. Benatti and R. Floreanini, Nucl. Phys. {\bf B488} (1997) 335
\smallskip
\item{14.} F. Benatti and R. Floreanini, Mod. Phys. Lett. {\bf A12} (1997) 1465
\smallskip
\item{15.} F. Benatti and R. Floreanini, Phys. Lett. {\bf B401} (1997) 337
\smallskip
\item{16.} F. Benatti and R. Floreanini, Completely positive dynamics of
correlated neutral kaons, Nucl. Phys. B, to appear
\smallskip
\item{17.} S. Hawking, Comm. Math. Phys. {\bf 87} (1983) 395
\smallskip
\item{18.} L.J. Garay, Spacetime foam as quantum thermal bath,
{\tt gr-qc/9801024}
\smallskip
\item{19.} C.B. Chiu, E.C.G. Sudarshan and G. Bhamathi, Phys. Rev. {\bf D46}
(1992) 3508
\smallskip
\item{20.} C. Bernardini, L. Maiani and M. Testa, Phys. Rev. Lett.
{\bf 71} (1993) 2687
\smallskip
\item{21.} I. Joichi, Sh. Matsumoto and M. Yoshimura, Time evolution
of unstable particle decay seen with finite resolution,
{\tt hep-ph/9711235}
\smallskip
\item{22.} Q. Wang and A.I. Sanda, Phys. Rev. D {\bf 55} (1997) 3131
\smallskip
\item{23.} L. Fonda, G.C. Ghirardi and T. Weber, Phys. Lett. 
{\bf B131} (1983) 309
\smallskip
\item{24.} P. Facchi and S. Pascazio, Temporal behaviour and quantum Zeno
region of an excited state of the hydrogen atom, University of Bari preprint, 
1997
\smallskip
\item{25.} E. Mihokova, S. Pascazio and L.S. Schulman, Hindered decay: 
quantum Zeno effect through electromagnetic field domination, 
Phys. Rev. {\bf A}, to appear
\smallskip
\item{26.} I.S. Gradshteyn and I.M. Ryzhik, {\it Tables of Integrals, Series
and Products} (Academic Press, San Diego, 1994)

\bye